\documentstyle[aps,prb,eqsecnum]{revtex} 
\newcommand{\beq}{\begin{equation}} 
\newcommand{\eeq}{\end{equation}} 
\newcommand{\nn}{\nonumber}

\newcommand{\jour}[4]{{\it #1}\ {\bf #2},\ #3\ (#4)}

\begin{document} 
 
 
\title{Fluctuation conductivity in superconductors\\ in strong electric fields} 
 
\author{Todor Mishonov,$^{a,b}$\thanks{$\mbox{Corresponding author:}$ 
$\mbox{phone: (+32 16) 327 183,}$~$\mbox{fax.: (+32 16) 327 983,}$~$\mbox{e-mail:}$~todor.mishonov@fys.kuleuven.ac.be} 
Anna Posazhennikova$^{a}$ and Joseph Indekeu$^{a}$ 
} 
 
\address{%
$^a$Laboratorium voor Vaste-Stoffysica en Magnetisme, 
 Katholieke Universiteit Leuven,\\ 
 Celestijnenlaan 200 D, B-3001 Leuven, Belgium 
} 
\address{ 
$^b$Department of Theoretical Physics, Faculty of Physics, Sofia University St. Kliment Ohridski,\\ 5 J. Bourchier 
Blvd., 1164 Sofia, Bulgaria } \preprint{Submitted to Phys. Rev. B; version of June 2, 2001} 
 
 
\maketitle 
 
\begin{abstract} 
We study the effect of a strong electric field on the fluctuation 
conductivity within the time-dependent Ginzburg-Landau theory for 
the case of arbitrary dimension. Our results are based on the 
analytical derivation of the velocity distribution law for the 
fluctuation Cooper pairs, from the Boltzmann equation. Special 
attention is drawn to the case of small nonlinearity of 
conductivity, which can be investigated experimentally. We obtain 
a general relation between the nonlinear conductivity and the 
temperature derivative of the linear Aslamazov-Larkin 
conductivity, applicable to any superconductor. For the important 
case of layered superconductors we derive an analogous relation 
between the small nonlinear correction for the conductivity and 
the fluctuational magnetoconductivity. On the basis of these 
relations we provide new experimental methods for determining both 
the lifetime constant of metastable Cooper pairs above $T_c$ and 
the coherence length. A systematic investigation of the 3rd 
harmonic of the electric field generated by a harmonic current can 
serve as an alternative method for the examination of the 
metastable Cooper-pair relaxation time. 
 
\end{abstract} 
 
 
\section{Introduction} 
\label{sec:introduction} 
Fluctuation phenomena in superconductors have been intensively 
studied for more than $3$ decades since the discovery of the 
fluctuation smearing of the superconducting transition. The 
interest in fluctuations was resurrected when the high-temperature 
superconductors (HTSC) were found. In conventional BCS-like 
superconductors the transition temperature $T_c$ marks a sharp 
dividing point between distinct regions of `superconducting' and 
`normal' behavior, and critical fluctuations are almost 
unobservable. Largely due to their high temperature, small 
coherence length and quasi-two-dimensional nature, HTSC show a 
significant smearing of the critical transition and thus the 
effect of fluctuations in the critical region of copper-oxides is 
much more pronounced and experimentally accessible. The 
conductivity, the specific heat, the diamagnetic susceptibility, 
the thermopower etc. have been observed to increase considerably 
in the normal state in the vicinity of the transition temperature 
in cuprate compounds. Importantly, for high-quality HTSC samples 
the interactions between fluctuations are negligibly small, 
fluctuations can be described as Gaussian ones and their theory is 
rather simple. As a result nowadays the study of fluctuation 
phenomena in superconductors occupies an essential and significant 
part of the whole physics of superconductivity. For a contemporary 
review we refer to the work of Larkin and 
Varlamov\cite{LarkinVarlamov}; cf. also the classical review by 
Skocpol and Tinkham\cite{SkocpolTinkham} and the recent review by 
Mishonov and Penev\cite{MishonovPenev} especially devoted to 
Gaussian fluctuations in layered superconductors. 
 
In this paper we are concerned with the effect of fluctuations of the superconducting order parameter on both the 
linear and the nonlinear conductivity near and above  the bulk superconducting critical temperature. In order to 
account for the fluctuation conductivity (or paraconductivity) one needs to apply the time-dependent 
generalization of the Ginzburg-Landau theory (TDGL). The theoretical study of fluctuation conductivity dates from 
the paper of Aslamazov and Larkin in 1968 based on the microscopic approach. Although a long time has elapsed, 
nonlinear effects have not been sufficiently investigated and the theoretical results obtained so far require 
specification. 
 
The aim of the present paper is to study the kinetics of the superconducting order parameter in a {\it strong 
electric field} within a simple phenomenological approach and to provide new experimental methods for determining 
such important material constants of superconductors as the lifetime of fluctuation Cooper pairs $\tau_0$ 
and the coherence length $\xi(0)$. 
 
We adopt the ideology of TDGL theory and show that the velocity 
distribution law of fluctuation Cooper pairs can be derived from a 
Boltzmann equation. The cases of superconducting wires, thin 
films, bulk superconductors and striped and layered 
superconducting systems are analyzed. We are particularly 
interested in the effects of nonlinearity of conductivity, which 
can be observed experimentally. We show that the strong critical 
behavior in the vicinity of $T_c$ of the third harmonic of the 
electric field generated by a harmonic current, can be used as an 
experimental technique for the determination of the metastable Cooper 
pair lifetime constant. 
 
This paper is organized as follows. In Section \ref{sec:boltzmann} we discuss the kinetic Boltzmann equation for 
fluctuation Cooper pairs in strong electric fields and present the momentum distribution law for Cooper pairs in 
the normal phase and the expression for the fluctuation current. Section \ref{sec:dimensionless} is mainly 
technical and devoted to dimensionless notations we introduce to make our calculation less complicated. The 
derivation of the time evolution of the distribution function using canonical variables is given in the Appendix. Section 
\ref{sec:conductiviy} explains the paraconductivity in layered materials, such as for example HTSC. 
We provide the analytical derivation of the conductivity correction in the normal phase in a Lawrence-Doniach 
superconductor. In Section \ref{sec:D-dimensional} the Aslamazov-Larkin conductivity dependent on the electric 
field is described in detail for one-, two- and three-dimensional superconductors. We study strong electric field 
effects on conductivity as well as weak electric field influences below the critical temperature. The interesting 
case of a striped superconductor, which is becoming very popular nowadays, and that of a thick film can be found 
in Section \ref{sec:thick}. The conductivity of such a superconductor depending on the thickness of the film, is 
obtained. The expressions for the nonlinear conductivity correction can be used for experimental data processing 
for extracting the lifetime constant of Cooper pairs, as is shown in Section \ref{sec:lifetime}. A new 
experimental technique for probing $\tau_0$, based on suggested measurements of the third harmonic of the electric 
field, can be found in Section \ref{sec:3rd}. Finally, a discussion and conclusions are given in the last section. 
 
\section{Boltzmann equation and formula for the current} 
\label{sec:boltzmann} \label{sec:current} 
 
As was already mentioned in Section \ref{sec:introduction}, the 
Ginzburg-Landau theory serves as an adequate tool for describing 
the fluctuation conductivity phenomena in superconductors. The 
important fundamental constant of the TDGL theory is $\tau_0$, 
which is proportional to the lifetime of metastable Cooper pairs 
in the normal state: 
\begin{equation} 
\label{} \tau(\epsilon)=\frac{\tau_0}{\epsilon}, 
\end{equation} 
where $\epsilon\equiv(T-T_c)/T_c$. 
 
Within the weak-coupling BSC theory in the case of negligible 
depairing mechanisms the temperature-independent constant $\tau_0$ 
satisfies the relation: 
\beq \tau_0^{\rm (BCS)}= \frac{\pi}{16}\frac{\hbar}{k_BT_c}.\eeq 
At present the experimental value for $\tau_0$, obtained for the 
layered cuprate superconductors, is in good agreement with the BCS 
theory\cite{MishonovPenev}. From an experimental point of view it is convenient to 
introduce the dimensionless ratio: 
\beq \tau_{\rm rel}= \frac{\tau_0}{\tau_0^{\rm (BCS)}}, \eeq 
which just characterizes the deviations of the experimental value from that which is theoretically derived from the 
microscopic theory for the case of small coupling and negligible depairing. Before establishing the relation 
between experimental results on fluctuation conductivity and the lifetime constant $\tau_0$ the important question 
to be clarified is what the momentum or velocity distribution of the metastable Cooper pair actually is. 
 
The charge of the Cooper pair is 
$\left|e^*\right|=2\left|e\right|$ and its effective mass $m^*$ is 
unambiguously experimentally accessible through many principally 
different methods: electrostatic charge modulation of the kinetic 
inductance\cite{inductance}, surface Hall effect\cite{Hall}, magnetoplasma waves and cyclotron 
resonance\cite{magnetoplasma}, Doppler effect of Cooper pair 
plasmons\cite{Doppler} and, which is most reliable, determination 
of the thermodynamic equilibrium electric potential related to the 
Bernoulli potential\cite{Bernoulli}. The velocity of a pair along 
the $x$-direction reads 
\beq 
v_x=\frac{p_x}{m_a^*} 
\eeq 
and along the $y$ and $z$ directions the velocities are 
correspondingly $v_y=p_y/m_b^*,$ and $v_z=p_z/m_c^*$. Note that we 
allow for the presence of mass anisotropy. 
 
It was shown \cite{MishonovDamianov,DamianovMishonov} within the general TDGL theory that, surprisingly, the 
momentum distribution law for the metastable Cooper pair is described by the classical Boltzmann equation 
introduced in physics in 1876 long before the electron was discovered. Nowadays this equation is discussed in many 
well-known textbooks on kinetics and solid state physics\cite{textbooks}. For a detailed description of the 
relaxation time approximation to the Boltzmann equation see also a well-known textbook\cite{Ashcroft}. The 
Boltzmann equation in our context is given by 
\beq \left(\frac{\partial}{\partial t} + e^* {\bf E}\cdot 
\frac{\partial}{\partial {\bf p}}\right)n({\bf 
p},t)= - \frac{n({\bf 
p},t)-\overline{n}_p}{\tau_p} = 
-\frac{n_p(t)}{\tau_p}+\frac{n_T}{\tau_0}. 
\label{Boltzmann} \eeq 
This kinetic equation for fluctuation Cooper pairs has been 
applied to the fluctuation Hall effect in thin superconducting 
films\cite{MishonovDamianov} and to the fluctuation paraconductivity 
within the framework of time-dependent Ginzburg-Landau 
theory\cite{DamianovMishonov}, cf. also Ref.\cite{LarkinVarlamov}. 
In Eq.(\ref{Boltzmann}) $n_p(t)=n({\bf p},t)$, and we introduce 
the dimensionless number $n_T=k_BT/a_0$, where 
\beq 
a_0=\frac{\hbar^2}{2m_a^*\xi_a^2(0)}=\frac{\hbar^2}{2m_b^*\xi_b^2(0)}=\frac{\hbar^2}{2m_c^*\xi_c^2(0)} 
\eeq 
is in fact proportional to the first coefficient in the Ginzburg-Landau expansion 
\beq 
a(\epsilon)= a_0\epsilon 
\eeq 
and 
\beq \overline{n}_p= \frac{k_BT}{\varepsilon_p-\mu} \eeq 
is a standard equilibrium distribution, where the energy spectrum 
of the layered superconductor with layer spacing $s$ is given by 
\beq \varepsilon_p= \frac{p_x^2}{2m_a^*} +\frac{p_y^2}{2m_b^*} 
+\frac{\hbar^2}{m_c^*s^2}\left[1-\cos\left(\frac{sp_z}{\hbar}\right)\right], 
\label{spectrum} \eeq 
and the chemical potential is defined through $ -\mu(T)= a(\epsilon)$. The momentum-dependent relaxation time 
$\tau_p$ obtained\cite{MishonovDamianov,DamianovMishonov} by TDGL theory reads as 
\begin{equation} 
\tau_p=\frac{\tau_0a_0}{\varepsilon_p+a(\epsilon)}. 
\end{equation}

Let us assume that the electric field is constant and applied 
along the $x$-direction in the $ab$-plane: ${\bf E}=(E_x,0,0)={\rm 
const}$. In this case from the Boltzmann equation 
(\ref{Boltzmann}) follows the stationary momentum distribution law 
\begin{eqnarray} 
n(p_x;\varepsilon_{\perp},E_x,T) &=& 
\frac{k_BT}{\hbar^2/2m_a^*\xi_a^2(0)} \int_{0}^{\infty} 
\exp\left\{-\left[\left(\frac{\xi_a(0)p_x}{\hbar}\right)^2 
+\frac{\varepsilon_{\perp}}{\hbar^2/2m_a^*\xi_a^2(0)}+\epsilon\right]\frac{t}{\tau_0} 
\right\} \nn\\ &&\times \exp\left\{ 
+\frac{e^*E_x\xi_a(0)}{\hbar/\tau_0}\,\frac{\xi_a(0)p_x}{\hbar}\left(\frac{t}{\tau_0}\right)^2 
-\frac{1}{3}\left(\frac{e^*E_x\tau_0}{\hbar/\xi_a(0)}\right)^2 
\left(\frac{t}{\tau_0}\right)^3\right\} \frac{dt}{\tau_0}, 
\label{npx} 
\label{Distribution} 
\end{eqnarray} 
where 
\beq \varepsilon_{\perp}(p_y,p_z)=\frac{p_y^2}{2m_b^*} 
+\frac{\hbar^2}{m_c^*s^2}\left[1-\cos\left(\frac{sp_z}{\hbar}\right)\right], 
\eeq 
and the physical meaning of the dummy parameter $t$ as time is 
revealed in the Appendix. For zero field  $E_x=0$ from (\ref{npx}) 
we naturally obtain 
\beq \overline{n}(p_x,\varepsilon_{\perp})=\frac{k_BT/a_0} 
{\left(\xi_a(0)p_x/\hbar\right)^2+\varepsilon_{\perp}/a_0+\epsilon}. 
\label{Equilibrium_Distribution} 
\eeq 

Once we have the momentum distribution law for the fluctuation 
Cooper pair, we can immediately proceed to the expression for the 
stationary fluctuation current for arbitrary dimension $D$: 
\beq {\bf j}=\sum_p e^*{\bf v}_p\,\frac{n_p}{\cal V} =\int e^*{\bf 
v}_p\,n({\bf p})\,\frac{d^Dp}{(2\pi\hbar)^D},\qquad 
{\bf v}_p=\frac{\partial \varepsilon_p}{\partial {\bf p}}, 
\eeq 
where ${\cal V}=L_xL_yL_z$ is the volume of a system. For the 
current along the $x$-direction in a layered superconductor we 
have 
\begin{eqnarray} 
j_x&=&e^* \frac{k_BT}{\hbar^2/2m_a^*\xi_a^2(0)} 
\int_{0}^{\infty}\frac{dt}{\tau_0} 
\exp\left\{-\frac{1}{3}\left(\frac{e^*E_x\tau_0}{\hbar/\xi_a(0)}\right)^2 
\left(\frac{t}{\tau_0}\right)^3\right\}\int_{-\pi\hbar/s}^{\pi\hbar/s}\frac{dp_z}{2\pi\hbar} 
\int_{-\infty}^{\infty}\frac{dp_y}{2\pi\hbar} \nn\\ &&\times 
\exp\left\{-\left[\epsilon+\left(\frac{p_y\xi_b(0)}{\hbar}\right)^2+ 
\left(\frac{2\xi_c(0)}{s}\sin\frac{sp_z}{2\hbar}\right)^2 
\right]\frac{t}{\tau_0}\right\} \nn\\ &&\times 
\int_{-\infty}^{\infty}\frac{p_x}{m_a^*} 
\left\{-\frac{t}{\tau_0}\left(\frac{\xi_a(0)p_x}{\hbar}\right)^2 
+\frac{e^*E_x\xi_a(0)}{\hbar/\tau_0}\left(\frac{t}{\tau_0}\right)^2\frac{\xi_a(0)p_x}{\hbar}\right\} 
\frac{dp_x}{2\pi\hbar}=\sigma_{xx}E_x. \label{current general} 
\end{eqnarray} 
A similar formula for the current was presented by 
Gor'kov\cite{Gor'kov} and modified for layered superconductors by 
Varlamov and Reggiani\cite{VarlamovReggiani}. 
 
\section{Dimensionless variables} 
\label{sec:dimensionless} 
 
To make our further calculations a bit less cumbersome we 
introduce a number of dimensionless variables: $k_x= 
\xi_a(0)p_x/\hbar$ and $k_y= \xi_b(0)p_y/\hbar \in 
(-\infty,\infty)$ are dimensionless momenta, $\theta= 
sp_z/\hbar\in(-\pi,\pi)$ is the Josephson phase and 
$u=t/\tau_0\in(0,\infty)$ is in fact the renormalized time. We 
also introduce 
\begin{equation} 
w\equiv\frac{\varepsilon_{\perp}}{a_0}= 
\left(\frac{p_y\xi_b(0)}{\hbar}\right)^2+ 
\left(\frac{2\xi_c(0)}{s}\sin\frac{sp_z}{2\hbar}\right)^2=k_y^2+\omega(\theta), 
\end{equation} 
where for the well-known Lawrence Doniach model 
$\omega(\theta)=\varepsilon_z(p_z)/a_0=\frac{1}{2}r(1-\cos\theta)=r\sin^2\frac{\theta}{2}$. 
The factor $r=\left(2\xi_c(0)/s\right)^2$ originates from the 
parametrization of the effective mass in $c$-direction, $m_c^*$, 
for an anisotropic GL model. 
 
It is convenient to deal with some dimensionless variable $f$, 
proportional to the electric field 
\beq f=\frac{e^*E_x\tau_0}{\hbar/\xi_a(0)},\eeq 
and with the parameter $g$, proportional to the nonlinear electric field correction 
\begin{equation} 
g=\frac{f^2}{12}=\frac{1}{12}\left(\frac{e^*E_x\xi_a(0)}{\hbar/\tau_0}\right)^2. 
\end{equation} 
With the above notations we can now rewrite our expression 
(\ref{npx}) for the momentum distribution of the metastable Cooper 
pair in the layered superconductor: 
\beq n(k_x;w,f,\epsilon)=n_T\int_0^\infty\exp\left[-\left(k_x^2+\epsilon+w\right)u + 
fk_xu^2-\frac{1}{3}f^2u^3\right]du. \label{n_dim} 
\eeq 
This function is a solution of the static Boltzmann equation which 
can be written in the form ($k=k_x$) 
\beq f\frac{d n(k)}{dk}=-\left(k^2+\epsilon+w\right)n(k)+n_T. \eeq 

Let us mention some details concerning the derivation of 
expression (\ref{n_dim}). Direct solution of the Boltzmann 
equation gives 
\begin{eqnarray} 
n(k)= \exp\left(-\frac{k^3/3+(\epsilon+w)k}{f}\right) 
\left[n_T\int_{k_0}^{k}\exp\left(\frac{\tilde 
k^3/3+(\epsilon+w)\tilde k}{f}\right)\frac{d\tilde 
k}{f}\right.\nn\\ 
\left.+\;n_0\exp\left(\frac{k_0^3/3+(\epsilon+w)k_0}{f}\right)\right]. 
\label{prom} 
\end{eqnarray} 
>From here it is easy to see that the boundary condition is 
$n(k_0)=n_0$. Then writing $\tilde k =k-u f$, introducing 
$u=(k-\tilde k)/f$ and using 
\beq 
\frac{\tilde k^3-k^3}{3f}= -k^2u+fku^2-\frac{1}{3}f^2u^3, 
\eeq 
we obtain from Eq.(\ref{prom}) 
\begin{eqnarray} 
n(k)&=&n_T\int_0^{(k-k_0)/f}\exp\left[-\left(k^2+\epsilon+w\right)u 
+ fku^2-\frac{1}{3}f^2u^3\right]du\nn\\ 
&&+n_0\exp\left(-\frac{k^3/3+(\epsilon+w)k}{f}\right)\exp\left(\frac{k_0^3/3+(\epsilon+w)k_0}{f}\right). 
\end{eqnarray} 
In the limit $k_0\rightarrow -\infty$ for $f>0$ or $k_0\rightarrow 
\infty$ for $f<0$  we have $n(-\infty)=n(\infty)=0=n_0$ and we 
arrive at Eq.(\ref{n_dim}). Finally the expression for the current 
Eq.~(\ref{current general}) in our notations for the layered 
superconductor takes the form 
\begin{eqnarray} 
j_x&=&e^* \frac{2k_BT}{\hbar}\xi_a(0) \int_{0}^{\infty}du\, 
\exp\left\{-\frac{1}{3}f^2u^3\right\}\int_{-\pi}^{\pi}\frac{d\theta}{2\pi 
s} \int_{-\infty}^{\infty}\frac{dk_y}{2\pi\xi_b(0)} \nn\\ 
&&\times\exp\left\{-\left[\epsilon+k_y^2 
+\frac{r}{2}(1-\cos\theta)\right]u\right\}\nn\\ &&\times 
\int_{-\infty}^{\infty}k_x \exp\left\{-uk_x^2+fu^2k_x\right\} 
\frac{dk_x}{2\pi\xi_a(0)}=\sigma_{xx}E_x. \label{current} 
\end{eqnarray} 
It is informative to compare this result with previously proposed 
expressions\cite{Gor'kov,VarlamovReggiani}. Our further analysis 
of the fluctuation conductivity is based on this general formula. 
 
\section{Paraconductivity in a layered metal} 
\label{sec:conductiviy} 
 
As we already pointed out in Section~\ref{sec:introduction} the 
fluctuations should be more important in high-temperature 
materials due in part to their high transition temperatures and 
extremely short coherence lengths (on the order of 10 \AA). Thus 
the study of Gaussian fluctuation effects on the conductivity in 
layered compounds is indispensable and of primary importance. In 
this section we examine in detail the electric field influence on 
paraconductivity in layered superconductors and derive 
analytically the correction to the conductivity in the normal 
phase. 
 
The integration over $k_x$ in the current expression (\ref{current}) can be easily performed: 
\beq \int_{-\infty}^{\infty}k_x 
\exp\left\{-uk_x^2+fu^2k_x\right\}dk_x =\frac{1}{2}\sqrt{\pi 
u}\,f\exp\left(\frac{1}{4}f^2u^3\right). \eeq 
Hence we have for the current in the layered superconductor the 
expression 
\begin{eqnarray} 
j_x&=& 
\frac{e^*k_BT}{2\pi^{1/2}\,\hbar}f 
\int_{-\pi}^{\pi}\frac{d\theta}{2\pi s} 
\int_{-\infty}^{\infty}\frac{dk_y}{2\pi\xi_b(0)} 
\int_{0}^{\infty}du\sqrt{u} 
\nn\\ 
&&\times\exp\left\{-\left[\epsilon+k_y^2 
+\frac{r}{2}(1-\cos\theta)\right]u-\frac{1}{12}f^2u^3\right\}=\sigma_{xx}E_x 
\end{eqnarray} 
and, consequently, the fluctuation conductivity reads 
\begin{eqnarray} 
\sigma_{xx}&=& \frac{2}{\pi^{1/2}} 
\frac{e^2k_BT\tau_0\xi_a(0)}{\hbar^2} 
\int_{-\pi}^{\pi}\frac{d\theta}{2\pi s} 
\int_{-\infty}^{\infty}\frac{dk_y}{2\pi\xi_b(0)} 
\int_{0}^{\infty}du\sqrt{u} \nn\\ 
&&\times\exp\left\{-\left[\epsilon+k_y^2 
+\omega(\theta)\right]u-gu^3\right\}=j_x/E_x. \label{ToBe1D} 
\end{eqnarray} 
After integrating over $k_y$ we obtain from (\ref{ToBe1D}): 
\beq \sigma_{xx}= 
\frac{e^2k_BT\tau_0\xi_a(0)}{\pi\hbar^2s\xi_b(0)} 
\int_{-\pi}^{\pi}\frac{d\theta}{2\pi} \int_{0}^{\infty}du 
\exp\left\{-\left[\epsilon+\omega(\theta)\right]u-gu^3\right\} 
=j_x/E_x. \label{ToBe2D} \eeq 

Thus the conductivity in the Lawrence-Doniach (LD) model can be 
expressed in the way 
\beq 
\sigma_{_{\rm LD}}= 
\frac{e^2k_BT\tau_0\xi_a(0)}{\pi\hbar^2s\,\xi_b(0)} 
\int_{0}^{\infty}du 
\exp\left\{-\left(\epsilon+\frac{r}{2}\right)u-gu^3\right\} 
\int_{-\pi}^{\pi} 
\exp\left\{\frac{r}{2}\cos(\theta)u\right\}\frac{d\theta}{2\pi} 
\eeq 
or, more conveniently, 
\beq \sigma_{_{\rm LD}}(\epsilon,g;r)= 
\frac{e^2k_BT\tau_0\xi_a(0)}{\pi\hbar^2s\,\xi_b(0)} 
\int_{0}^{\infty}du 
\,\exp\left\{-\left(\epsilon+r/2\right)u-gu^3\right\} I_0\left(r 
u/2\right), \label{ToBe3D} \eeq 
where 
\beq I_0(x)=\frac{1}{\pi}\int_0^\pi{\rm e}^{x\cos\theta}d\theta 
=\sum_{n=0}^\infty\left[\frac{(x/2)^n}{n!}\right]^2 
=J_0(ix) 
\eeq 
is the Bessel function for imaginary argument. 
 
In the case of zero electric field  $E_x=0$ the integral over $u$ 
in Eq.(\ref{ToBe3D}) is easy to compute since 
\beq 
\int_{-\pi}^{\pi}\frac{d\theta}{2\pi} \int_{0}^{\infty}du 
\exp\left\{-\left[\epsilon+\frac{r}{2}\left(1-\cos\theta\right)\right]u\right\} 
=\int_{-\pi}^{\pi}\frac{d\theta}{2\pi} \frac{1} 
{\left(\epsilon+\frac{r}{2}\right)-\frac{r}{2}\cos\theta} 
=\frac{1} {\sqrt{\epsilon(\epsilon+r)}}, \label{integral} 
\eeq 
and we obtain the Lawrence-Doniach formula for the fluctuation 
conductivity of a layered superconductor 
\beq \sigma_{_{\rm LD}}(\epsilon;r)= 
\frac{e^2k_BT\tau_0\xi_a(0)}{\pi\hbar^2s\,\xi_b(0)} 
\frac{1}{\sqrt{\epsilon(\epsilon+r)}}. \eeq 
Hereafter we define, for simplicity, 
$\sigma(\epsilon;r)\equiv\sigma(\epsilon,0;r)$. 
 
 It is simpler to deal with the dimensionless 
function (cf. Varlamov and Reggiani\cite{VarlamovReggiani} 
Eq.~(7)) 
\beq \varsigma_{_{\rm LD}}(\epsilon,g;r) =\frac{\sigma_{_{\rm 
LD}}(\epsilon,g;r)}{\sigma_{_{\rm LD}}(\epsilon;r)} 
=\sqrt{\epsilon(\epsilon+r)} 
\int_{0}^{\infty}\exp\left\{-\left(\epsilon+r/2\right)u-gu^3\right\}I_0\left(r 
u/2\right)du. \label{sigma} \eeq 
Then Eq.~(\ref{ToBe3D}) reads 
\beq \sigma_{_{\rm LD}}(\epsilon,g;r)=\sigma_{_{\rm 
LD}}(\epsilon;r)\,\varsigma_{_{\rm 
LD}}(\epsilon,g;r).\label{sigmaeps} \eeq 

It is important to note that the product of two exponents in 
(\ref{sigma}) can be expressed in terms of the third-derivative 
operator acting on the function of $\epsilon$ 
\beq 
\exp\left(-gu^3\right) \exp\left(-\epsilon u\right) 
=\exp\left(g\frac{\partial^3}{\partial\epsilon^3}\right) 
\exp\left(-\epsilon u\right). 
\eeq 
This key observation immediately leads us to the simple general 
relation between the nonlinear fluctuation conductivity and the 
linear one in evanescent field, $\sigma(\epsilon)$, 
\beq 
\sigma(\epsilon>0,g)=\exp\left(g\frac{\partial^3}{\partial\epsilon^3}\right)\sigma(\epsilon),\qquad 
\label{deriv} 
\eeq 
which is similar to the relation for the magnetoconductivity, 
derived in Ref.\cite{MishonovPenev}. For our dimensionless function 
(\ref{sigma}) we get 
\begin{equation} 
\varsigma(\epsilon,g)=\frac{1}{\sigma(\epsilon)} 
\exp\left(g\frac{\partial^3}{\partial\epsilon^3} 
\right)\sigma(\epsilon). 
\end{equation} 

In the limit $g\rightarrow 0$ the electric-field-dependent 
conductivity can be written in a simple form 
\begin{equation} 
\sigma(\epsilon,f)=\sigma(\epsilon)+\Delta\sigma_f(\epsilon,f) 
\approx\left(1+g\frac{\partial^3}{\partial\epsilon^3}\right)\sigma(\epsilon) 
\label{DiffEqn}, 
\end{equation} 
and thus the nonlinear electric-field correction to the 
conductivity reads 
\begin{equation} 
\Delta\sigma_f(\epsilon,f)\approx\frac{f^2}{12}\frac{\partial^3}{\partial\epsilon^3}\sigma(\epsilon). 
\end{equation} 
 
Now we can derive the conductivity, dependent on the electric 
field in the Lawrence-Doniach model of a layered superconductor. 
In principle we may start either from the right-hand-side of 
Eq.(\ref{integral}) (i.e., we can take the third derivative after 
averaging over the Josephson phase), or we can perform the 
differentiation before integrating over $\theta$. The latter is a 
useful method for the evaluation of complicated integrals 
necessary for experimental data processing and leads to 
\beq 
\oint\frac{d\theta}{2\pi}\frac{1}{\left[\epsilon+\frac{r}{2}\left(1-\cos\theta\right)\right]^4} 
=-\frac{1}{6}\frac{\partial^3}{\partial\epsilon^3}\frac{1}{\sqrt{\epsilon(\epsilon+r)}} 
=\frac{\epsilon^3+\frac{3}{2}r\epsilon^2+\frac{9}{8}r^2\epsilon+\frac{5}{16}r^3} 
{\left[\epsilon(\epsilon+r)\right]^{7/2}}. \eeq 
Finally for the Lawrence-Doniach conductivity we have 
\beq \sigma_{_{\rm LD}}(\epsilon,g;r) \approx \left[ 
1-\frac{1}{2}\frac{\epsilon^3+\frac{3}{2}r\epsilon^2+\frac{9}{8}r^2\epsilon+\frac{5}{16}r^3}{(\epsilon+r)^3} 
f_\epsilon^2\right] \sigma_{_{\rm LD}}(\epsilon;r), \qquad 
f_\epsilon \ll 1, \eeq 
where 
\beq \qquad 
f_\epsilon=\frac{|e^*E_x|\xi_a(\epsilon)\tau(\epsilon)}{\hbar}=\frac{|f|}{\epsilon^{3/2}}=\frac{|E_x|}{E_c(\epsilon)},\qquad 
g_\epsilon=\frac{1}{12}f_\epsilon^2=\frac{g}{\epsilon^3}, \eeq 
and other notations are 
\beq 
E_c(\epsilon)=\frac{\hbar}{|e^*|\xi(\epsilon)\tau(\epsilon)}=E_c(0)\epsilon^{3/2}, 
\qquad 
E_c(0)=\frac{\hbar}{|e^*|\xi(0)\tau_0},\label{electric_field} \eeq 
\beq \xi_{a,b,c}(\epsilon)=\frac{\xi_{a,b,c}(0)}{\epsilon^{1/2}}, 
\qquad 
r_\epsilon=\frac{r}{\epsilon}=\left(\frac{2\xi_c(\epsilon)}{s}\right)^2. 
\eeq 
Our dimensionless $\varsigma$ function can now be rewritten in the 
scaled variables, as can be seen from (\ref{sigma}), 
\beq 
\varsigma_{_{\rm LD}}(\epsilon,g;r) =\varsigma_{_{\rm 
LD}}(1,g_\epsilon;r_\epsilon). 
\eeq 
In conclusion of this section we obtain the formula for the 
electric-field-dependent correction to the conductivity in the 
Lawrence-Doniach model for a layered superconductor: 
\beq \Delta\sigma_f(\epsilon,E_x) 
=-\frac{4k_BTe^4\left[\xi_a(0)\tau_0\right]^3}{\pi\hbar^4s\xi_b(0)}\, 
\frac{\left[\epsilon^3+\frac{3}{2}r\epsilon^2+\frac{9}{8}r^2\epsilon+\frac{5}{16}r^3\right]} 
{\left[\epsilon(\epsilon+r)\right]^{7/2}}E_x^2 \equiv\Delta 
j_x/E_x. \label{cubicLD} \eeq 
%
 
\section{Aslamazov-Larkin conductivity for D-dimensional superconductors} 
\label{sec:D-dimensional} 
 
In this section we are concerned with the derivation of the 
Aslamazov-Larkin electric-field-dependent conductivity for bulk 
superconductors ($D=3$), thin films ($D=2$) and wires (D=1). The 
results can be generalized for the case of arbitrary dimension. 
 
We start with a one-dimensional superconductor. In order to derive 
the expression for the one-dimensional fluctuation conductivity we 
should set $p_y=p_z=0$ in (\ref{spectrum}) and exclude from 
Eq.~(\ref{ToBe1D}) the integration over the perpendicular 
component of the momenta 
\beq 
{\bf 
p}_{\perp}=(p_y,p_z)=\left(\frac{\hbar 
k_y}{\xi_b(0)},\frac{\hbar\theta}{s}\right), 
\eeq 
(the electric field is as usual parallel to the x-direction). 
We immediately get the final expression 
\begin{eqnarray} 
\sigma_{\rm 1D}= \frac{2}{\pi^{1/2}} 
\frac{e^2k_BT\tau_0\xi_a(0)}{\hbar^2} \int_{0}^{\infty}\sqrt{u} 
\exp\left\{-\epsilon u-gu^3\right\}du=\frac{I_x}{E_x}. \label{1D} 
\end{eqnarray} 
For the derivation of a two-dimensional conductivity, induced by 
the fluctuations, we have to cancel averaging with respect of the 
Josephson phase  $\theta$ in Eq.~(\ref{ToBe2D}) and we find 
\beq 
\sigma_{\rm 2D}= 
\frac{e^2k_BT\tau_0\xi_a(0)}{\pi\hbar^2\xi_b(0)} \int_{0}^{\infty} 
\exp\left\{-\epsilon u-gu^3\right\}du =\frac{j_x^{\rm (2D)}}{E_x}. 
\label{2D} 
\eeq 

The case of a layered superconductor turns into the bulk (3D) one 
in the limit $r\rightarrow\infty$, when the distance between 
the layers tends to zero. Taking into account the asymptotics for 
the Bessel function in Eq.~(\ref{ToBe3D}), cf. 
Ref.~\cite{VarlamovReggiani}: 
\begin{equation} 
I_0(x\gg1)\approx\frac{{\rm e}^x}{\sqrt{2\pi x}},\qquad 
\exp\left(-\frac{r}{2}u\right)I_0\left(\frac{r}{2}u\right)\approx\frac{1}{\sqrt{\pi 
r u}}, 
\end{equation} 
we obtain the bulk fluctuation conductivity expression 
\beq \sigma_{\rm 3D}= 
\frac{e^2k_BT\tau_0\xi_a(0)}{2\pi^{3/2}\hbar^2\xi_b(0)\xi_c(0)} 
\int_{0}^{\infty} \exp\left\{-\epsilon 
u-gu^3\right\}\frac{du}{\sqrt{u}} =\frac{j_x^{\rm (3D)}}{E_x}. 
\label{3D} \eeq 

These results bring us naturally to the general expression for the 
fluctuation conductivity for arbitrary dimension $D$ and $T$ above 
$T_c$ : 
\beq \sigma_{\rm 
D}(\epsilon,g)=\frac{e^2k_BT\tau_0\xi_a^2(0)}{2^{D-2}\pi^{D/2}\hbar^2\xi^D(0)} 
\int_{0}^{\infty} \exp\left\{-\epsilon 
u-gu^3\right\}\frac{du}{u^{(D-2)/2}}, \label{general} 
\label{sigmaD} \eeq 
(cf. Dorsey\cite{Dorsey} (1991), Eq.~(47), for $\epsilon>0$). 
 
One can express Eq.(\ref{general}) in a form similar to 
Eq.~(\ref{sigmaeps}): 
\beq \sigma_{\rm D}(\epsilon,f)=\sigma_{\rm 
D}(\epsilon)\varsigma_{\rm D}(g_\epsilon), \eeq where 
\begin{equation} \sigma_{\rm 
D}(\epsilon)=\frac{4\Gamma\left(\frac{4-D}{2}\right)}{\left(4\pi\right)^{D/2}} 
\frac{e^2}{\hbar^2}k_BT\frac{\tau(\epsilon)\xi_a^2(\epsilon)}{\xi^D(\epsilon)}, 
\eeq 
is the fluctuation conductivity in zero electric field, and 
\begin{equation} 
\xi^D(\epsilon)\equiv \left\{\begin{array}{ll} 
 \xi_a(\epsilon)\quad \mbox{for} \quad D=1, \\ 
   \xi_a(\epsilon)\xi_b(\epsilon) \quad \mbox{for} \quad D=2,\\ 
   \xi_a(\epsilon)\xi_b(\epsilon)\xi_c(\epsilon) \quad \mbox{for} \quad D=3,\\ 
 \mbox{...} 
\end{array} \right. 
\end{equation} 
The function 
\begin{equation} 
\varsigma_{\rm 
D}(g_\epsilon)=\frac{1}{\Gamma\left(\frac{4-D}{2}\right)} 
\int_{0}^{\infty}v^{1-D/2}{\rm e}^{-v-g_\epsilon v^3}d v,\qquad 
v=\epsilon u 
\eeq 
is a dimensionless function containing the electric-field 
dependence of the fluctuation conductivity.

For small electric field $f_\epsilon\ll 1$, in accordance with 
Eq~(\ref{DiffEqn}), we have: 
\beq \varsigma_{\rm D}(f_\epsilon)\approx 
1-\left(\frac{4-D}{2}\right)\left(\frac{6-D}{2}\right)\left(\frac{8-D}{2}\right)\frac{f^2_\epsilon}{12}. 
\eeq 
%

\subsection{Strong electric field expansion} 
\label{sec:strong} 
 
In this subsection we study the conductivity analytically in the limit of electric fields that are large compared 
with the reduced temperature distance to the critical point. To this end we introduce the scaled variable 
$\epsilon_f=12^{1/3}\,\epsilon/f^{2/3}$ and focus on the regime of small $\epsilon_f$. The sign of $\epsilon_f$ is 
arbitrary. Our analysis applies to $T \le T_c$ as well as $T>T_c.$ Changing the integration variable $u$ in 
Eq.~(\ref{sigmaD}) to $f^{2/3}u$ allows us to extract the dominant field dependence at the critical point. 
Subsequently we use the series expansion of $\exp(-\epsilon u)$ and integrate it term by term for obtaining a 
series in $\epsilon_f$. We arrive at the result 
\beq 
J_{\rm D}(\epsilon,g)\equiv \int_{0}^{\infty} 
\exp\left\{-\epsilon u-gu^3\right\}\frac{du}{u^{(D-2)/2}}= 4 
\frac{f^{-(4-D)/3}}{12^{(D+2)/6}}\sum_{n=0}^{\infty} 
\frac{(-\epsilon_f)^n}{n!}\Gamma\left(\frac{2n+4-D}{6}\right). 
\label{Jintegral} 
\eeq 
The convergence is very good for $\left|\epsilon_f \right|\leq 1$. For instance, for $\epsilon_f = 0.2$ the errors 
in zeroth, first, second and third order are, respectively, $12\%, 1\%, 0.07\%$ and 40 ppm (parts per million). 
This series leads to the following expression for the conductivity 
\beq \sigma_{\rm D}(\epsilon,f)=\sigma_{\rm D}(f)\Sigma_{\rm 
D}\left(\epsilon_f\right) \eeq with \beq \qquad \Sigma_{\rm 
D}(\epsilon_f)\equiv 1+\sum_{n=1}^{\infty} 
\frac{(-\epsilon_f)^n}{n!}\frac{\Gamma\left(\frac{2n+4-D}{6}\right)}{\Gamma\left(\frac{4-D}{6}\right)} 
\eeq 
and 
\begin{eqnarray} 
\sigma_{\rm D}(f)&=&\frac 
{\Gamma\left(\frac{4-D}{6}\right)e^2k_BT\tau_0\xi_a^2(0)f^{-(4-D)/3}} 
{3^{(D+2)/6}2^{(4D-10)/3}\pi^{D/2}\hbar^2\xi^D(0)}\nn\\ &=&\frac 
{4\Gamma\left(\frac{4-D}{6}\right)k_BT\tau_0^{(D-1)/3}} 
{\left[2\sqrt{\pi}\xi(0)\right]^DE_x^{(4-D)/3}} 
\left(\frac{e\xi_a(0)}{\sqrt{3}\hbar} \right)^{(D+2)/3} 
\end{eqnarray} 
is the fluctuation conductivity at $T_c$; 
cf.\cite{VarlamovReggiani,Hurault,Dorsey}. 
 
\subsection{Weak electric fields below $T_c$} 
\label{sec:low} 
 
In order to study the fluctuation conductivity for temperatures slightly below $T_c$ and in small electric fields 
we must take into account that the limit of zero field is singular due to the occurrence of bulk 
superconductivity. The contribution to the conductivity that we calculate in this subsection must be well 
separated physically from that due to the onset of bulk order. The approximation scheme we develop here is a good 
one for relatively weak electric fields, which satisfy at $T<T_c$ the condition 
\beq 
f_\epsilon\equiv\frac{|f|}{|\epsilon|^{3/2}} 
=\frac{2\left|eE_x\right|\xi_a(0)\tau_0}{\hbar|\epsilon|^{3/2}}<1. 
\eeq 
Especially below $T_c$ we have to take into account that the 
temperature distance to the critical point $\epsilon$ is 
renormalized as can be seen in self-consistent mean-field-like 
approximations which we will briefly consider later. The origin of 
this effect lies in the non-linear character of the TDGL equation, 
which cannot be neglected for high densities of fluctuation Cooper 
pairs. The selfconsistent approximation decouples the 
non-linearity, resulting in a linear problem with a modified 
$\epsilon$, to be denoted by $\epsilon_r$. To alleviate the 
notation we will postpone this substitution until the end of this 
subsection. 
 
For $T < T_c$ we write $\epsilon = -\left| \epsilon \right|$ and to the integral in Eq.~(\ref{Jintegral}) we apply 
the Gaussian saddle point approximation for weak electric fields. This amounts to looking for the maximum of the 
argument of the exponential function, since the remaining factor is an algebraic function of $u$ and therefore 
slowly varying. Defining 
\beq F(u)=-|\epsilon|u+\frac{f^2u^3}{12}\eeq 
the saddle-point approximation can be written as 
\beq {\rm e}^{-F(u)} \approx {\rm e}^{-F(u_{_0})} 
\sqrt{\frac{2\pi}{F^{\prime\prime}(u_0)}}\,\delta(u-u_0), 
\label{joseph} 
\eeq 
where $u_0$ is the minimum of $F(u)$, given by 
\beq u_0=2\frac{\sqrt{|\epsilon}|}{|f|}; 
\label{u_0} 
\eeq 
the time interval $t_0=\tau_0u_0$ has a transparent physical 
meaning, $\left|e^*E_x\right|t_0=2\hbar/\xi_a(\epsilon).$ In order 
for this approximation to be accurate, the condition 
\beq F^{\prime\prime}(u_0)u_0^2 \gg 1 \label{Gauss_condition} \eeq 
must be satisfied, which is 
equivalent to 
\beq {f_\epsilon} \ll 1 .\eeq 
This condition can be seen to arise from two requirements. Firstly, already in $D=2$, in the absence of the power 
of $u$ in the integral, the validity of the saddle-point approximation requires \beq \exp 
\left(-\frac{1}{2}F^{\prime\prime}(u_0)u_0^2 \right ) \ll 1 \eeq in order for the integration interval to be 
extendable to $(-\infty, \infty)$. Secondly, in the presence of slowly varying additional factors in the 
integrand, we must check the consistency of the approximation by performing a Taylor expansion of the algebraic 
function of $u$ about $u_0$. If we denote this function by $G(u)$, we may approximate this by the constant 
$G(u_0)$ in the integral, provided $G(u)$ deviates only weakly from linearity in a neighbourhood of width $w_F$ 
around $u_0$, where $w_F$ is the standard deviation of the Gaussian function. This is fulfilled when 
\beq G^{\prime\prime}(u_0) w_F^2/G(u_0) \ll 1 .\eeq 
Since $w_F^2 =1/F^{\prime\prime}(u_0)$ and $G(u)$ is simply a power of $u$, this condition coincides with 
Eq.~(\ref{Gauss_condition}). 
 
Within the range of validity of the saddle-point approximation we thus arrive at the following result for the 
integral 
\beq J_{\rm D}(-|\epsilon|,f)\approx\sqrt{\pi}\,|\epsilon |^{-(4-D)/2} 
\left(\frac{2}{f_\epsilon}\right)^{(3-D)/2} 
\exp\left(\frac{4}{3f_\epsilon}\right),\qquad{\rm for}\quad{\rm 
e}^{2/f_\epsilon}\gg1. 
\eeq 
According  to Eq.~(\ref{sigmaD}) and taking into account the renormalization of $\epsilon$ to $\epsilon_r$ we 
obtain 
\beq 
\sigma_{\rm D}(-|\epsilon|,f)\approx\frac{e^2k_BT}{2^{(3D-7)/2}\pi^{(D-1)/2}\hbar^2} 
\frac{\tau(\epsilon_r)\xi_a^2(\epsilon_r)}{\xi^D(\epsilon_r)}f_\epsilon^{(D-3)/2} 
\exp\left(\frac{4}{3f_\epsilon}\right). \eeq 
The replacement $\epsilon\rightarrow\epsilon_r$ derives from the use of a Maxwell-type selfconsistent approach (cf. 
Ref.\cite{MishonovPenev}) for solving the Boltzmann equation. The implicit equation relating the renormalized and 
bare parameters is given by 
\beq 
\epsilon_r-\epsilon=\frac{\mu_0}{m_{ab}^*}\left[e^*\lambda_{ab}(0)\right]^2 
\int \frac{d^Dp}{(2\pi\hbar)^D}n({\bf p},{\bf E},\epsilon_r), 
\eeq 
with 
\beq \frac{1}{\lambda_{ab}^2(0)}\equiv 
-T_c\left.\frac{d}{dT}\frac{1}{\lambda_{ab}^2(T)}\right|_{T_c^-}, 
\label{selfconsistent} 
\eeq 
where details of the procedure of ultraviolet regularization will be presented elsewhere. 
 
The most significant aspect of our result is the dramatic exponential increase in the fluctuation conductivity for 
small electric fields. Therefore, for $f_\epsilon\ll 1$, the fluctuation part becomes of the same order of 
magnitude as the normal-state background, $\sigma_D(-|\epsilon|,f)\simeq \sigma_N$, and the fluctuation 
conductivity will no longer be just a perturbation but a significant part of the total conductivity 
\beq \sigma_{\rm tot}=\sigma_D(\epsilon,f)+\sigma_N(T)= 
j_{\rm tot}/E_x. 
\eeq 

This extraordinary increase of the conductivity naturally leads to a {\em minimum} in the current as a function of 
the applied field, in agreement with the Gor'kov analysis\cite{Gor'kov} that the current-voltage characteristic 
must have a perfectly noticeable section corresponding to negative differential conductivity. At a suitable 
potential difference between the ends of the film, the generation of radiation should be observed. Indeed, for 
every $\epsilon<0$ the generation of radiation will start for electric fields lower than the critical one, $E_{\rm 
gen}(\epsilon=-|\epsilon|)$, determined by the criterion 
\beq \left.\frac{dj_{\rm tot}(E_x)}{d E_x}\right|_{E_{\rm gen}}=0, 
\eeq 
which is equivalent to 
\beq \qquad f_{\rm 
gen}\left.\frac{d}{d f}\sigma(-|\epsilon|,f)\right|_{f_{\rm 
gen}}+\sigma(-|\epsilon|,f_{\rm gen}) =-\sigma_N(T). 
\eeq 

Within the selfconsistent approximation Eq.~(\ref{selfconsistent}) it is easy to obtain theoretical formulae for 
the case when the fluctuations are nonlinear 
\beq 
E_{\rm gen}(-|\epsilon_r|)= E_c(0) 
f_{\rm gen}(-|\epsilon_r|), 
\eeq 
where the unit of electric field $E_c(0)$ is defined in 
Eq.~(\ref{electric_field}), and this will be a nontrivial test of 
the validity of the selfconsistent approximation applied to the 
TDGL equation and, following from it, the Boltzmann equation for 
fluctuation Cooper pairs. Now we are addressing cases important 
for the applications, in which the fluctuation superconductivity 
can be easily investigated. 
 
\section{Striped superconductors and thick films} 
\label{sec:thick} 
 
The latest achievements in nanotechnology provide us in principle with a tool for a practical realization of so 
called striped superconductors with controlled parameters. For this it is necessary to cut stripes from a superconducting 
film using some appropriate lithographic technology. The amazing observation about the striped materials is that 
they are of ``intermediate" dimensionality, i.e., they are neither 1-dimensional systems nor 2-dimensional ones. 
The closer the striped superconductor is to its critical temperature $T_c,$ the more ``perfectly" 
two-dimensional material it becomes, because the stripes become increasingly coherent. Analogously to the 
Lawrence-Doniach model for the layered superconductor we can describe this situation in terms of ``dimensional 
crossover". In the present section we show that the fluctuation conductivity longitudinal to the stripes can be 
derived following the standard procedure for layered superconductors. We have also to mention that probably some 
underdoped cuprates are naturally striped and this phenomenon has been at the center of the attention attracted by HTSC 
during the last few years. 
 
It is obvious from  Section~\ref{sec:D-dimensional}, for example, 
that the LD conductivity for a layered superconductor can be 
naturally derived from a two-dimensional AL conductivity just by 
integrating that over the momentum in the direction perpendicular 
to the plane 
\beq 
\sigma_{_{\rm LD}}(\epsilon,f)=\int\frac{dp_z}{2\pi\hbar} 
\sigma_{\rm 
2D}\left(\epsilon+\frac{\varepsilon_z(p_z)}{a_0},f\right) 
=\int_{-\pi}^{\pi}\sigma_{\rm 
2D}\left(\epsilon+\frac{r}{2}(1-\cos\theta),f\right)\frac{d\theta}{2\pi 
s}. 
\eeq 
Thus the fluctuation conductivity for a striped superconductor 
reads 
\beq \sigma_{\rm striped}(\epsilon,f)= \frac{1}{s}\oint\sigma_{\rm 
1D}\left(\epsilon+\frac{r}{2}(1-\cos\theta),f\right)\frac{d\theta}{2\pi}, 
\eeq 
where $s$ now stands for the period of the stripes. 
 
For a thick film with thickness $d_{\rm film}$ we have to sum over 
the discrete spectrum of the energy associated with the motion in 
z-direction: 
\beq \varepsilon_z(p_z)=\frac{p_z^2}{2m^*_c},\qquad 
\frac{\varepsilon_z(p_z)}{a_0}=\left(\frac{\pi\xi_c(0)}{d_{\rm 
film}}\right)^2n_z^2,\eeq 
since 
\beq p_z=\frac{\pi\hbar}{d_{\rm film}}n_z,\qquad 
n_z=0,1,2,3,\dots . \eeq 
In the expression for the fluctuation conductivity of a thick film 
we have now summation instead of integration 
\beq \sigma_{\rm film}(\epsilon,f)=\frac{1}{d_{\rm 
film}}\sum_{p_z} \sigma_{\rm 
2D}\left(\epsilon+\varepsilon_z(p_z)/a_0,f\right)= \frac{1}{d_{\rm 
film}}\sum_{n_z}^\infty \sigma_{\rm 
2D}\left(\epsilon+\left(\pi\xi_c(0)/d_{\rm 
film}\right)^2n_z^2,f\right), \eeq 
which for zero electric field can be readily performed using 
\beq \sum_{n=0}^\infty\frac{1}{a^2+b^2n^2}=\frac{\pi}{2ab\tanh 
\left(\frac{\pi a}{b}\right)}+\frac{1}{2a^2}. \eeq 
Finally we obtain the expression for the 3D conductivity of the 
film, which interpolates between 2D and 3D behaviour analogously 
to the conductivity for a layered superconductor: 
\begin{eqnarray} 
\sigma_{\rm film}(\epsilon)&=& 
\sigma_{2D}(\epsilon)\left\{\frac{1}{2d_{\rm film 
}}+\frac{1}{2\xi_c(\epsilon)\tanh\left(\frac{d_{\rm film}}{\xi_c(\epsilon)}\right)}\right\} 
=\frac{e^2\tau_{rel}}{16\hbar\epsilon}\left\{\frac{1}{2d_{\rm film}}+ 
\frac{\sqrt{\epsilon}\,\coth\left(\frac{d_{\rm film}\sqrt{\epsilon}}{\xi_c(0)}\right)}{2\xi_c(0)}\right\}, 
\end{eqnarray} 
\beq 
\sigma_{LD}(\epsilon)= 
\frac{\sigma_{\rm 2D} 
(\epsilon)}{\sqrt{s^2+\left(2\xi_c(\epsilon)\right)^2}}= 
\frac{e^2\tau_{\rm rel}}{16\hbar\sqrt{\epsilon^2s^2+\epsilon\left(2\xi_c(0)\right)^2}}. 
\eeq 
Note that the thick film becomes ``rightly" {\it two}-dimensional in the vicinity of $T_c$, thus the 
dimensionality of it decreases, whereas for the layered superconductor the dimensionality goes up from $D=2$ to 
$D=3$ as we approach the critical temperature. In such a way a thick film of a strongly anisotropic layered 
superconductor, $\xi_c(0)\ll s \ll d_{\rm film},$ can have two\cite{VarlamovYu} dimensional crossovers.

\section{Determination of the lifetime constant $\tau_0$} 
\label{sec:lifetime} 
 
In the current section we show that conductivity measurements in a 
strong electric field can serve as a method for probing 
fundamental properties of superconductors such as the lifetime 
constant of metastable Cooper pairs $\tau_0$ and the coherence 
length $\xi(0)$. We demonstrate that our theoretical results can 
be effectively used for experimental data processing and 
determination of both $\tau_0$ and $\xi(0)$. 
 
As a rule in an experiment the temperature dependence of a 
resistivity $\rho_{\rm exp}(T)$ is examined. The experimentally 
measured conductivity is consequently $\sigma_{\rm 
exp}=1/\rho_{\rm exp}(T)$. The in-plane current, if we take into 
account the first nonlinear correction, can be written in the 
general form 
\beq j_x=\sigma_{\rm exp}E_x-{\cal A}(\epsilon)E_x^3, \eeq 
since the nonlinear correction to the fluctuation conductivity is 
$\Delta\sigma_f=-{\cal A}E_x^2$. 
 
Of primary interest for us are the superconducting films. Let us 
consider the two-dimensional BCS-like superconductor. In this case 
the parameter $r$ that determines the effective dimensionality of 
the superconductor is zero, and the periodicity of the 
Lawrence-Doniach model $s\rightarrow d_{\rm film}$ ($d_{\rm film}$ 
is the thickness of the superconducting film). Thus the 
two-dimensional current is $j_x^{\rm (2D)}=d_{\rm film}j_x$. 
 
>From the expression for the LD fluctuation conductivity 
Eq~(\ref{cubicLD}) it follows: 
\beq j_x^{(\rm 
2D)}=\left(\sigma_N(T)+\frac{e^2}{16\hbar}\frac{\tau_{\rm 
rel}}{\epsilon}\right)E_x 
-\frac{4\pi^2}{\hbar}\left[\xi_{ab}(0)/k_BT\right]^2\frac{e^4\tau_{\rm 
rel}^3}{\left(8\epsilon\right)^4}E_x^3. \eeq 

In order to study the fluctuation effect on conductivity one 
should plot first of all the paraconductivity contribution to the 
resistance $1/(1/\rho_{\rm exp}(T)-1/\rho_N(T))$ as a function of 
$T$ (see for example \cite{FioryHebardGlaberson}, where 
indium/oxide films were examined). For InO$_{\rm x}$ films the 
value of $\tau_{\rm rel}= 1.16$ can be used as a tool for the 
determination of the in-plane coherence length $\xi_{ab}(0)$. 
 
In the general case of a layered LD superconductor the coefficient 
in the nonlinear correction to fluctuation conductivity according 
to Eq.~(\ref{cubicLD}) reads 
\beq {\cal 
A}(\epsilon)=\frac{4k_BTe^4\left[\xi_a(0)\tau_0\right]^3}{\pi\hbar^4s\xi_b(0)}\, 
\frac{\left[\epsilon^3+\frac{3}{2}r\epsilon^2+\frac{9}{8}r^2\epsilon+\frac{5}{16}r^3\right]} 
{\left[\epsilon(\epsilon+r)\right]^{7/2}}. \eeq 
In this case we have three coherence lengths $\xi_a(0),$ $\xi_b(0),$ $\xi_c(0),$ and the current anisotropy 
$J_{\rm max}/J_{\rm min}$. 
 
In this paper we develop a model-free method for the determination 
of the lifetime of Cooper pairs $\tau_0$ and the coherence length 
from the experimental results for the fluctuation conductivity. 
The final result is derived on the basis of the 
Eq.~(\ref{DiffEqn}) after three-fold integration over some time 
interval ($\epsilon_1,\epsilon_2$). We obtain the following 
expression for the lifetime 
\beq \tau_0=\sqrt{\frac{3}{2}}\,\frac{\hbar}{|eE_x|\xi_a(0)} 
\left\{\frac{\int_{\epsilon_1}^{\epsilon_2}\left(\tilde\epsilon-\epsilon_1\right)^2 
\left[-\Delta\sigma_f(\tilde\epsilon,f)\right]d\tilde\epsilon} 
{\sigma(\epsilon_1)-\left[\sigma(\epsilon_2) 
+\left(\epsilon_1-\epsilon_2\right)\sigma^{\prime}(\epsilon_2) 
+\frac{1}{2}\left(\epsilon_1-\epsilon_2\right)^{2}\sigma^{\prime\prime}(\epsilon_2)\right]} 
\right\}^{1/2} \label{tau0}. \eeq 

In (\ref{tau0}) we have the notations 
\beq 
\sigma^{\prime}(\epsilon)=\frac{\partial}{\partial\epsilon}\sigma(\epsilon),\qquad 
\sigma^{\prime\prime}(\epsilon)=\frac{\partial^2}{\partial\epsilon^2}\sigma(\epsilon),\qquad 
\eeq 
and $\epsilon_2\simeq 0.2$ determines the upper bound on the temperature below which the fluctuation conductivity 
can be reliably measured. In order to simplify the formula (\ref{tau0}) we can make the approximation 
$\epsilon_2\rightarrow\infty$. Since the fluctuation phenomena are negligible already for $T-T_c\approx 0.15T_c$ 
the following asymptotic conditions can be imposed in the limit $\epsilon_2\rightarrow\infty,$ 
\beq 
\sigma(\epsilon_2)\rightarrow 0,\qquad 
\epsilon_2\sigma^{\prime}(\epsilon_2)\rightarrow 0,\qquad 
\epsilon_2^{2}\sigma^{\prime\prime}(\epsilon_2)\rightarrow 0. 
\eeq 
 
Finally we obtain the formulae for the lifetime of metastable 
Cooper pairs and the in-plane coherence length, which can be 
applied for experimental data processing: 
\begin{eqnarray} 
\tau_0&=&\sqrt{\frac{3}{2}}\,\frac{\hbar}{|eE_x|\xi_a(0)} 
\left\{\frac{1}{\sigma(\epsilon)} 
\int_{\epsilon}^{\infty}\left(\tilde\epsilon-\epsilon\right)^2 
\left[-\Delta\sigma_f(\tilde\epsilon,f)\right]d\tilde\epsilon 
\right\}^{1/2}, \label{tau} \\ 
\tau_{\rm rel}&=&\sqrt{\frac{3}{2}}\,\frac{16k_BT}{\pi 
|eE_x|\xi_a(0)} \left\{\frac{1}{\sigma(\epsilon)} 
\int_{\epsilon}^{\infty}\left(\tilde\epsilon-\epsilon\right)^2 
\left[-\Delta\sigma_f(\tilde\epsilon,f)\right]d\tilde\epsilon 
\right\}^{1/2}, \label{taurel} \\ 
\xi_{ab}(0)&=&\sqrt{\frac{\Phi_0}{\pi |B_z|}} 
\left\{\frac{1}{\sigma(\epsilon)} 
\int_{\epsilon}^{\infty}\left(\tilde\epsilon-\epsilon\right) 
\left[-\Delta\sigma_h(\tilde\epsilon,h)\right]d\tilde\epsilon 
\right\}^{1/4} 
\end{eqnarray} 
cf. Ref.\cite{MishonovPenev} Eqs.~(201-203). 
Here \beq 
-\Delta\sigma_h(\epsilon,h)=\frac{h^2}{4}\frac{\partial^2}{\partial\epsilon^2}\sigma(\epsilon),\qquad 
h\ll \epsilon \eeq 
is a nonlinear correction to the magnetoconductivity 
and the magnetic field \beq h=\frac{B_z}{B_{c2}(0)}\eeq is 
oriented perpendicular to the $ab$-plane, where 
\beq 
B_{c2}(0)=\left.-T_c\frac{dB_{c2}(T)}{dT}\right|_{T_c} 
=\frac{\Phi_0}{2\pi\xi_{ab}^2(0)} 
\eeq 
is the slope of the upper 
critical field and 
\beq 
\Phi_0=\frac{2\pi\hbar}{|e^*|} 
\eeq 
is the magnetic flux quantum. 
 
\section{Conductivity correction by detection of 3rd harmonics} 
\label{sec:3rd} 
As an alternative method for probing the fundamental constants of 
the BCS theory  we suggest the systematic investigation of the 
third harmonic of the electric field generated by a harmonic 
current. Third-harmonic measurements are easier to perform than 
those of resistivity, and, moreover, the effect arising from 
fluctuations is exceptionally pronounced. 
 
In general $AC$ current response and investigation of higher harmonics is a standard method for investigation of 
nonlinear effects on superconductivity. A homogeneous electric field 
\beq E_x(t)=E_0\cos(\omega t), \eeq 
for example, creates a small nonlinear response for the first 
harmonic and a cubic field dependence of the 3rd harmonic of the 
current 
\begin{eqnarray} 
j_x(t)&=&\sigma E_x(t)-{\cal A}E_x^3(t)\nn\\ &=&j_{1f}\cos(\omega 
t)+j_{3f}\cos(3\omega t), \label{j3f} 
\end{eqnarray} 
where for the amplitudes of the harmonics we have 
\beq j_{1f}/E_0=\sigma-\frac{3}{4}{\cal A}E_0^2,\qquad 
j_{3f}=-\frac{1}{4}{\cal A}E_0^3.  \eeq 
If necessary, a smooth analytical normal part of the nonlinear 
coefficient ${\cal A}$ can be subtracted from the experimental 
data ${\cal A}_{\rm exp}$ 
\beq {\cal A}_{\rm exp}={\cal A}_N(T)+{\cal A}(\epsilon),\qquad 
{\cal A}_N(T)=A+BT+CT^2, \eeq 
in order to extract pure fluctuation behaviour from the nonlinear 
coefficient of the conductivity correction 
\beq \Delta\sigma_f=-{\cal A}(\epsilon)E_x^2. \eeq 
If we use the so defined conductivity correction $\Delta\sigma_f$ 
the electric field $E_x$ is actually cancelled in Eq.~(\ref{tau}) 
and we have to use the coefficient ${\cal A}(\epsilon)$ in the 
expression for $\tau_0$, i.e. 
\beq \tau_0=\sqrt{\frac{3}{2}}\,\frac{\hbar}{|e|\xi_a(0)} 
\left\{\frac{1}{\sigma(\epsilon)} 
\int_{\epsilon}^{\infty}\left(\tilde\epsilon-\epsilon\right)^2 
{\cal A}(\tilde\epsilon) d\tilde\epsilon \right\}^{1/2}. 
\label{tauA} \eeq 
There is no doubt that the electric field is a useful tool for a 
theoretical analysis but for the experimental realization of the 
suggested method we have to apply a harmonic current and to 
measure the harmonics of the voltage 
\beq I(t)=I_0\cos\omega t,\qquad U(t)=U_{1f}\cos\omega 
t+U_{3f}\cos3\omega t+U_{5f}\cos5\omega t+\dots . \eeq 
For small current amplitudes used to avoid heating of the sample 
the voltage response is in first approximation linear and we have 
Ohm's law for the resistance of the superconductor strip with 
length $L,$ width $w$ and thickness $d_{\rm film}$, 
\beq U_{1f}=R(T)I_0,\quad R(T)=\rho(T)\frac{L}{wd_{\rm 
film}},\quad \rho(T)=\frac{1}{\sigma_{\rm exp}(T)},\quad 
E_0=\frac{U_{1f}}{L}=\frac{\rho(T)I_0}{wd_{\rm film}}. \eeq 
Then the absence of the 3rd harmonic of the current, $j_{3f}=0$, 
according to Eq.~(\ref{j3f}) with $E_x(t)=U(t)/L$ gives 
\beq \frac{U_{3f}}{\rho(T)L}\approx\frac{1}{4}{\cal 
A}(\epsilon)\left(\frac{U_{1f}}{L}\right)^3 \eeq 
to lowest order in $I_0$, and finally we obtain 
\beq {\cal 
A}(\epsilon)\approx4\frac{L^2}{\rho(T)}\frac{U_{3f}}{(U_{1f})^3} 
=4\frac{L^3}{wd_{\rm film}}\frac{I_0U_{3f}}{(U_{1f})^4}. \eeq 
In this way the nonlinear coefficient necessary for the 
determination of the lifetime constant in Eq.~(\ref{tauA}) can be 
expressed through the electronically measured current $I_0,$ 
voltage amplitudes $U_{3f}$ and $U_{1f}$, and the geometrical 
parameters of the strip $L$, $w$ and $d_{\rm film}.$ So the 
suggested experiment can be performed in every laboratory involved 
in investigations of superconductivity. 
 
Let us describe qualitatively the temperature dependence of the intensity of the 3rd harmonic when the temperature 
is increased. In the superconducting state the voltage response is negligible and the 3f signal will appear abruptly when we 
reach the critical temperature. After a sharp maximum at $T_c$ the 3f signal will decrease with smaller slope and 
in the normal region the 3f signal will be small again and created only by the 2f oscillations of the temperature and 
the temperature dependence of the resistivity. In short we predict that $U_{3f}$ will have a $\lambda$-shaped asymmetric 
critical singularity. The location of this $\lambda$-point provides a new method for the determination of the critical 
temperature of superconductors based on the properties of fluctuation phenomena. Our selfconsistent theoretical 
calculation is applicable above $T_c$ where $U_{3f}$ is much smaller than the value at the $\lambda$-point, but still 
clearly detectable experimentally. 
 
\section{Discussion and conclusions} 
Let us start  analyzing the results derived with the momentum 
distribution Eqs.~(\ref{Distribution}) and 
(\ref{Equilibrium_Distribution}). We have a characteristic 
velocity related to the equilibrium distribution 
\beq 
v_c(\epsilon)=\frac{\hbar}{m_a^*\xi_a(\epsilon)}=v_c(0)\sqrt{\epsilon}, 
\qquad m_a^*v_c(0)=\frac{\hbar}{\xi_a(0)} \eeq 
and Eq.~(\ref{Equilibrium_Distribution}) can now be rewritten as 
\beq n(v_x,\varepsilon_\perp)=\frac{n_T}{\left[v_x/ 
v_c(\epsilon)\right]^2+\epsilon+\varepsilon_\perp/a_0}. \eeq 
According to the Drude consideration, a small electric field creates a drift velocity 
\beq v_{\rm drift}(\epsilon)= 
\frac{e^*E_x\tau(\epsilon)}{m_a^*}=\frac{e^*E_x\tau_0}{m_a^*\,\epsilon} 
\eeq 
and the dimensionless electric field is just the ratio of those 
two velocities, 
\beq f_\epsilon=\frac{|v_{\rm 
drift}(\epsilon)|}{v_c(\epsilon)},\qquad 
v_c(\epsilon)\simeq\sqrt{\epsilon}\,v_{\rm 
Fermi}\exp\left(-\frac{1}{\rho_{\rm Fermi}V_{\rm pairing}}\right). 
\eeq 
>From a microscopic point of view the characteristic thermal velocity $v_c(\epsilon)$ is proportional to the Fermi 
velocity times the small parameter of the BCS theory, the famous exponent which contains the density  of states at 
the Fermi level and the matrix element of the pairing interaction. This order of magnitude estimation is 
applicable to anisotropic gaps as well. In addition we have a critical slowing down multiplier $\sqrt{\epsilon}$. 
Those two factors significantly decrease the characteristic velocity and make possible the experimental 
observation of the electric-field correction to the fluctuation conductivity proportional to $f_\epsilon^2.$ For 
normal metals the ratio $v_{\rm drift}^2/v_{\rm Fermi}^2\ll1$ is extremely small and only AC oscillations of the 
temperature $T(t)$, mainly 2f, can create harmonics in the voltage response. This effect should also be carefully 
taken into account for cuprate films for which the thermal resistance between the substrate and the film can be very 
high; this will be the subject of another work. 
 
Let us also consider in short the $\omega \tau$-quasiparticle criterion to check whether or not Cooper pairs are 
quasiparticles in the usual sense of condensed matter physics, i.e., $\tau_p\left(\epsilon_p-\mu\right)/\hbar\gg1.$ For 
${\bf p}=0$ taking the microscopic value for the lifetime we have 
\beq 
\tau(\epsilon)\frac{a(\epsilon)}{\hbar}=\tau_0^{\rm (BCS)}\,\frac{a_0}{\hbar} 
=\frac{\pi}{16}\,\frac{a_0}{k_BT_c}=\frac{\pi}{16}\,\frac{1}{n_T}\ll 1. 
\eeq 
This strong inequality means that fluctuation Cooper pairs are not quasiparticles. The notion of Cooper pairs is only a language 
to describe the properties of slowly decaying diffusion modes of the superconducting order parameter above $T_c$. 
Analogously the ``mean free path" $l(\epsilon)=v(\epsilon)\tau(\epsilon)$ is also very short relative to the 
correlation radius: 
\beq 
\frac{l(\epsilon)}{\xi(\epsilon)}=\frac{v(\epsilon)\tau(\epsilon)}{\xi(\epsilon)} 
=\frac{\hbar}{m^*\xi(\epsilon)}\,\frac{\tau(\epsilon)}{\xi(\epsilon)}=2\frac{a_0\tau_0}{\hbar} 
=\frac{\pi}{8}\,\frac{1}{n_T}\ll1. 
\eeq 
As an illustration let us take a set of parameters corresponding to a high-$T_c$ cuprate: 
$T_c=90$~K, $k_BT_c=7.76$~meV, $m_{ab}^*=11 m_e,$ where $m_e$ is the mass of a free electron, 
$\xi_{ab}(0)=11$~\AA. Then $a_0=\hbar^2/2m_{ab}^*\xi_{ab}^2(0)=2.86$~meV and $n_T=k_BT_c/a_0= 2.71>1.$ 
The last inequality ensures that $n(\epsilon)=n_T/\epsilon\gg1$ and this condition of applicability of 
Rayleigh-Jeans statistics justifies the treatment of the Ginzburg-Landau $\Psi$-function as a classical complex 
field. 
 
In view of the new effects which can be predicted using the 
derived velocity distribution we consider the possibility of 
supercooling of the normal phase, cf. Ref.\cite{Gor'kov}, to be 
very interesting. In this case the fluctuation conductivity could 
create a negative differential conductivity, which opens 
perspectives for many technical applications. In order to prevent 
the nucleation of superconductivity from regions where the current 
densities and electric fields are very small, depairing impurities 
should be introduced in the contact area of the microbridge. This 
could be realized, for example, by evaporation of Ni on the wide 
area of the Al microbridge or by Mn ions in cuprate films. In both 
cases the central narrow region of the microbridge should be 
protected. In a sample prepared under these conditions the 
criterion of negative differential conductance can be easily 
satisfied and so the predicted generation of oscillations would 
probably be the best example of significant fluctuation effects in 
superconductors. 
 
Whether the threshold for the generation regime can be described 
within the nonlinear theory or whether we need to calculate the 
fluctuation density and the renormalized temperature 
$T_r=T_c(\epsilon_r+1)$ in a selfconsistent way, depends on the 
numerical value of the Ginzburg number. In any case the analyzed 
solution of the Boltzmann equation suffices to predict a 
cross-over from positive to negative differential conductivity, as 
the field decreases. As a precursor of oscillations, when the 
electric field is decreased, due to strong nonohmic behaviour and 
low dissipation, the sample will be an excellent frequency mixer. 
The incipient bulk conductivity should always be taken into 
account, because the supercooled normal state is metastable and 
applying a voltage to the superconducting state leads as a rule to 
a space- and time-inhomogeneous phase. 
 
Some words should be added concerning the history of the kinetic equation introduced in 1876 by Boltzmann. This 
was the first use of probability concepts in a dynamical theory but the real recognition of the Boltzmann equation 
was stimulated by the electronic industry in the second half of the 20th century. Indeed, the Boltzmann equation 
is an essential tool for understanding how electronic devices work. 
 
In sum, in our paper we have used the Boltzmann equation not only as a didactical instrument but also as a means 
of deriving new results. The Boltzmann equation for fluctuation Cooper pairs is a consequence of the 
time-dependent Ginzburg-Landau equation. This equation has been derived from microscopic theory and in this sense the 
Boltzmann equation is a tool for the application of the microscopic theory of superconductivity. That is why the 
Boltzmann equation can be used to predict the results of new experiments, to help in their interpretation, and 
even to correct some previously obtained results derived in the frame-work of microscopic theory.

\acknowledgments 
 
We highly appreciate the interesting discussion of our paper with A.~Varlamov and his stimulating comments. This 
research has been supported by the Belgian DWTC, IUAP, the Flemish GOA and VIS/97/01 Programmes. A.I.P. is KUL 
Junior Fellow (F/99/045) and T.M.M. is KUL Senior Fellow (F/00/038). 
 
 
\section*{Appendix} 
\label{sec:appendix} 
In the Boltzmann equation Eq.~(\ref{Boltzmann}) the decay rate of 
fluctuation Cooper pairs depends on the kinetic energy 
$\varepsilon(\bf p).$ As an illustration here we will analyze the 
isotropic GL model with 
\beq \frac{1}{\tau({\bf 
p})}=\frac{1}{\tau_0}\frac{\varepsilon({\bf 
p})+a_0\epsilon}{a_0},\quad 
\varepsilon({\bf p})=\frac{{\bf p}^2}{2m^*},\quad 
{\bf p}= m^*{\bf v}, \quad 
{\bf v}=\frac{\partial \varepsilon({\bf p})}{\partial {\bf p}}. 
\eeq 
The Boltzmann equation is a dynamic equation and we consider it 
natural to trace the time evolution of the velocity distribution 
starting from some arbitrary initial distribution 
\beq n({\bf p}, t=0)=n_0({\bf p}). \eeq 
As a rule dynamic problems can be more elegantly analyzed using 
canonical variables. That is why we will analyze the momentum 
distribution $N({\bf P},t)$ defined with respect to the canonical 
momentum 
\beq {\bf P}= {\bf p}+e^*{\bf A}. \eeq 
For a constant and space-homogeneous electric field ${\bf E}={\rm 
const}$ in the $\varphi=0$ gauge we have 
\begin{eqnarray} 
{\bf A}(t)=-t{\bf E},\;\; 
{\bf A}(t=0)=0,\; 
{\bf P}({\bf p},t)\equiv{\bf p}-e^*{\bf E}t,\; 
\left(\frac{\partial {\bf P}}{\partial {\bf p}}\right)_t=\openone_{D\times D},\; 
\left(\frac{\partial {\bf P}}{\partial t}\right)_p=-e^*{\bf E}. 
\end{eqnarray} 
Now we will solve the Boltzmann equation using the distribution 
$N({\bf P},t)$. The physical quantity is the same -- the number of 
Cooper pairs living at some momentum point -- but mathematically 
the functions are different and their correspondence is given by 
\beq N({\bf P},t)\equiv n({\bf p},t). \eeq 
After this change of the variables the sum of partial derivatives 
on the left-hand-side of the Boltzmann equation reduces to a usual 
derivative 
\begin{eqnarray} 
\left(\frac{\partial}{\partial t}+e^*{\bf 
E}\cdot\frac{\partial}{\partial {\bf p}}\right)N({\bf P}({\bf p 
},t),t)&=& \frac{\partial N}{\partial t}+\frac{\partial 
N}{\partial{\bf P}}\cdot\frac{\partial {\bf P}}{\partial t} 
+\frac{\partial N}{\partial {\bf P}}\cdot\frac{\partial {\bf 
P}}{\partial {\bf p}} \cdot e^*{\bf E}\nn\\ &=&\frac{\partial 
N({\bf P},t)}{\partial t} =\frac{d N_P(t)}{d t}. 
\end{eqnarray} 
The physical interpretation is very simple, in an external electromagnetic field the canonical momentum is 
conserved and reduces simply to a label with which the distribution can be parametrized. In this derivation it is 
essential to note that when a partial derivative is taken with respect to one argument, the other argument of 
the function is kept constant. As a consequence, $\left(\partial n/\partial t\right)_p\neq \left(\partial 
N/\partial t\right)_P$. Taking into account that the decay rate and the equilibrium distribution are functions of 
the kinetic momentum ${\bf p}={\bf P}-e^*{\bf A},$ the Boltzmann equation takes the form 
\beq \frac{d N_P(t)}{d t}=-\frac{N_P(t)-\overline n({\bf 
P}-e^*{\bf A})}{\tau({\bf P}-e^*{\bf A})} 
=-\left[\frac{\left({\bf 
P}+e^*{\bf E}t\right)^2}{2m^*}+a_0\epsilon\right]\frac{N_P(t)}{a_0\tau_0} 
+\frac{n_T}{\tau_0}. \eeq 
Introducing now dimensionless time and dimensionless canonical 
momentum 
\beq \tilde u =\frac{t}{\tau_0},\quad {\bf q}=\xi(0){\bf P}/\hbar, 
\quad q=k-f\tilde u, \eeq 
we find that in the 1D case the Boltzmann equation takes the convenient form 
\beq \frac{d N_P(\tilde u)}{d \tilde u} =-\left[\left(q+f\tilde 
u\right)^2+\epsilon\right]N_P(\tilde u)+n_T. \eeq 
One can easily check that the function 
\begin{eqnarray} 
N_q(\tilde u)&=&n_T\exp\left\{-\frac{1}{3f}(q+f\tilde 
u)^3-\epsilon \tilde u\right\} \int_0^{\tilde 
u}\exp\left\{\frac{1}{3f}\left(q+fu^\prime\right)^3+\epsilon 
u^\prime\right\}du^\prime\nn\\ 
&+&n_0(q)\exp\left\{-\frac{1}{3f}\left[\left(q+f\tilde 
u\right)^3-q^3\right]-\epsilon \tilde u\right\} 
\label{N_q} 
\end{eqnarray} 
satisfies this ordinary linear differential equation with the 
initial condition 
\beq N_q(\tilde u=0)=n_0(q). \eeq 
Now introducing a new dimensionless variable 
\beq u\equiv\tilde u-u^\prime \eeq 
we find for the distribution $n_k(\tilde u)\equiv N_q(\tilde u)$ 
with respect to the kinetic momentum $k=q+f\tilde u$, 
\begin{eqnarray} 
n_k(\tilde u) &=&n_T\int_0^{\tilde 
u}\exp\left\{-\left(k^2+\epsilon\right)u+kfu^2-\frac{1}{3}f^2u^3\right\}du 
\nn\\ &+& n_0(k-f\tilde 
u)\exp\left\{-\left(k^2+\epsilon\right)\tilde u+kf\tilde 
u^2-\frac{1}{3}f^2\tilde u^3\right\}. 
\label{n_k} 
\end{eqnarray} 
Note that the argument of $n_0$ is a conserved quantity $k-f\tilde u$, 
cf. also Eq.~(\ref{N_q}). 
 
The transition to the limit $t=\tau_0\tilde u\rightarrow\infty$ is 
very simple. After several relaxation times we have the stationary 
distribution 
\beq n(k)\equiv n_k(\tilde u\rightarrow\infty)= 
n_T\int_0^\infty\exp\left\{-\left(k^2+\epsilon\right)u+kfu^2-\frac{1}{3}f^2u^3\right\}du. 
\eeq 
Restoring $k\rightarrow k_x,$ $\epsilon\rightarrow \epsilon+w,$ 
and $n(k)\rightarrow n(k_x;w,f,\epsilon)$ we return to the 
solution of the static Boltzmann equation Eq.~(\ref{n_dim}). 
Consequently the dummy parameter $t=\tau_0u$ in 
Eq.~(\ref{Distribution}) has the meaning of the time interval 
between the birth of fluctuation Cooper pairs and the moment when 
we measure the current. Hence the integrand in the formula for the 
current Eq.~(\ref{current general}) has the meaning of the part of 
the current given by Cooper pairs born a time span $t$ \textit{before} 
the moment of the measurement. We consider it interesting that 
according to Eqs.~(\ref{joseph}, \ref{u_0}) this ``age distribution" has a sharp 
$\delta$-like maximum at $t_0=\tau_0u_0$ for the supercooled 
normal phase in small electric fields. 
 
Finally, it is amusing to note that although the fluctuation Cooper pairs are neither particles nor 
quasi-particles the conservation of canonical momentum implies Newton's equation of motion 
\begin{equation} 
\label{cond-mat psw: xxxx} 
\frac{d{\bf p}}{dt}=e^*{\bf E}. 
\label{manuscript number: xxx} 
\end{equation} 
%
 
 
%

\begin{references} 
%
\bibitem{LarkinVarlamov} 
 A.~Larkin and A.~Varlamov, 
 \textit{Fluctuation Phenomena in Superconductors}, Sec.4.3 in 
 ``Physics of Conventional and Unconventional Superconductors", 
  Edited by K.~Bennemann and J.~B~Ketterson, 
  (Springer, Berlin, to appear in 2002). 
%
\bibitem{SkocpolTinkham} 
 W.~J.~Skocpol and M.~Tinkham, 
 \jour{Rep. Prog. Phys.}{38}{1094}{1975}, Sec.~5.3, Fig.~15, and references therein. 
%
\bibitem{MishonovPenev} 
 T.~Mishonov and E.~Penev, 
 \jour{International Journal of Modern Physics}{B14}{3831-3878}{2000}. 
%
\bibitem{textbooks} 
 L.~D.~Landau and E.~M.~Lifshitz, 
  {\it Physical Kinetics}, (Pergamon, New York, 1973) Sec.25., Eq.~(25.12); 
 A.~A.~Abrikosov, {\it Fundamentals of the Theory of Metals}, (North Holland, Amsterdam, 1988), 
 J.~M.~Ziman, {\it Principles of the Theory of Solids}, (Cambridge Univ Press, 1972) Chap.7; J.~M.~Ziman, 
 {\it Electrons and Phonons}, (Oxford Univ Press, 1960), 
 I.~M.~Lifshitz, M.~I.~Kaganov, \jour{Sov. Phys. 
 Usp.}{2}{831}{1960}, \jour{\text{ibid}}{5}{878}{1963}, \jour{\text{ibid}}{8}{805}{1966} 
 (\jour{Usp. Fiz. Nauk}{69}{419}{1959}, 
 \jour{\text{ibid}}{78}{411}{1962}, \jour{\text{ibid}}{87}{389}{1965}). 
%
\bibitem{Ashcroft} N.~W.~Ashcroft, N.~D.~Mermin, {\it Solid State 
Physics}, (Holt, Rinehart and Winston, New York, 1976); Chap.16. 
%
\bibitem{inductance} 
 T.~M.~Mishonov, 
 \jour{Phys. Rev. Lett.}{67}{3195}{1991}; 
 A.~T.~Fiory, A.~F.~Hebard, R.~H.~Eick, P.~M.~Mankiewich, R.~E.~Howard, M.~L.~O'Malley, 
 \jour{\textit{ibid.}}{67}{3196}{1991}. 
%
\bibitem{Hall} 
T.~Mishonov and N.~Zahariev, 
\jour{Superlattices and Microstructures}{26}{57-60}{1999}. 
%
\bibitem{magnetoplasma} 
 T.~M.~Mishonov, 
 \jour{Phys. Rev. B}{42}{6715}{1990}; 
 K.~Karrai, E.~Choi, F.~Dunmore, S.~Liu, X.~Ying, Qi Li, T.~Venkatesan, H.~D.~Drew, Qi Li, D.~B.~Fenner, 
  \jour{Phys. Rev. Lett.}{69}{355}{1992}. 
%
\bibitem{Doppler} 
 T.~M.~Mishonov, 
 \jour{Phys. Rev. B}{50}{4004}{1994}. 
%
\bibitem{Bernoulli} 
 T.~M.~Mishonov, 
 \jour{Phys. Rev. B}{50}{4009}{1994}. 
%
\bibitem{MishonovDamianov} 
 T.~M.~Mishonov and D.~Ch.~Damianov, 
 \jour{Czech. J. Phys.}{46}{(Suppl. S2) 631}{1996}. 
%
\bibitem{DamianovMishonov} 
 D.~Ch.~Damianov and T.~M.~Mishonov, 
 \jour{Superlattices and Microstructures}{21}{467}{1997}. 
%
\bibitem{Gor'kov} 
 L.~P.~Gor'kov, 
 \jour{Zh. Eksp. Teor. Fiz. Pis. Red.}{11}{52-56}{1970}; 
 (Engl. transl. \jour{Sov. Phys.-JETP Lett.}{11}{32-35}{1970}); Eq.~(5). 
%
\bibitem{VarlamovReggiani} 
 A.~A.~Varlamov and L.~Reggiani, 
 \jour{Phys. Rev. B}{45}{1060-1063}{1992}; Eqs.~(1,6). 
%
\bibitem{Hurault} 
 J.~P.~Hurault, 
 \jour{Phys. Rev.}{179}{494-496}{1969}. 
%
\bibitem{Dorsey} 
 A.~Dorsey, 
 \jour{Phys. Rev. B}{43}{7575-7585}{1991}. 
%
\bibitem{VarlamovYu} A.~A.~Varlamov, L.~Yu, 
\jour{Phys. Rev. B}{44}{7078-7080}{1991}. 
 
%
\bibitem{FioryHebardGlaberson} A.~T.~Fiory, A.~F.~Hebard and W.~I.~Glaberson, 
 \jour{Phys. Rev. B}{28}{5075-5087}{1983}, Fig.~3. 
%
\end{references}
\end{document}